\documentclass{ragtime} % NO optional arguments are required
\title[Stress-energy tensor of a radiating sphere inclosing black hole]%
      {Stress-energy tensor of a radiating sphere inclosing black hole}

\author[M. Wielgus and M. Abramowicz]%    Now let's start the paper title authors:
       {Maciek Wielgus\at[]{1,2,a}
       and Marek Abramowicz\at[]{3,1,4,b}\\% Termination of authors' block; if
        \ins{1}Nicolaus Copernicus Astronomical Center,
        Polish Academy of Sciences,\splitins[1]% This is how to break an
        ul. Bartycka 18, 00-716 Warszawa, Poland\\% Termination of the first affiliation.
        \ins{2}Institute of Micromechanics and Photonics,  ul. A. Boboli 8, 02-525 Warszawa, Poland\\% Termination of the second affiliation.
         \ins{3}Institute of Physics, Faculty of Philosophy and Science, Silesian University in Opava,\splitins[1]
         Bezru\u{c}ovo n\'{a}m. 13, 746 01 Opava, Czech Republic\\% Termination of the first affiliation.
        \ins{4}Physics Department, Gothenburg University, 412-96 G\"oteborg, Sweden\\% Termination of the second affiliation.    
                \ins{a}\Email{maciek.wielgus@gmail.com}\\
                 \ins{b}\Email{marek.abramowicz@physics.gu.se}} % This is how to present E-mail.

\coentry{Z. Stuchl\'{\i}k, J. Kov\'a\v{r} and J. Cimrman}%
       {Equilibrium of spinning test particles in equatorial plane\par
        of Kerr--de~Sitter spacetimes}

\newcommand{\be}{\begin{equation}}
\newcommand{\ee}{\end{equation}}

\begin{document}

% Citation of references in abstract should generally be avoided to
% ensure self-consistency of the abstract.  If you do insist on citation(s)
% within the abstract, you should use the \bibentry command, which forces
% the _complete_bibliographic_entry_ to appear in the abstract.
% With the `nonatbib' optional class argument this feature is not available.
\begin{abstract}
We consider a uniformly luminous radiating sphere and a static black hole located in the center of that sphere. We give analytic formulas for radiation stress-energy tensor components in such a configuration, for the observer located at an arbitrary distance from the static black hole horizon.
\end{abstract}

% The key words are to be separated by the N-dash surrounded by spaces,
% the left one non-breakable.  The name concatenation like Kerr--de~Sitter
% should also be typed with N-dash but with no spaces around (compare,
% e.g., Levi-Civita, which is a single person):

% It is good to provide as many as \label's possible, but never start the
% key with a numeral.  This makes problems with pdflatex processing.
\section{Introduction}\label{intro}%%%%%%%%%%%%%%%%%%%%%%%%%%%%%%%%%%%%%%%%

In \citet{Abramowicz1990} an analytic formula for the stress-energy tensor of uniformly radiating static relativistic star was found, for ZAMO (zero angular momentum observer) at the arbitrary distance from the star surface. The result is very useful for investigations of the test particle motion in a~curved spacetime, under the radiation four-force influence and several groups pursue this branch of research in recent years, see, e.g.,  \citet{Bini2010}, \citet{Oh2011}, \citet{Stahl2012},  \citet{Wielgus2012}, \citet{Stahl2013}. Here we consider the extended problem, i.e., the observer may be located inside the uniformly radiating luminous sphere of radius $R$. In that way we complete the solution of \citet{Abramowicz1990}, allowing for any ratio of the radiating sphere's and observer's radii in the Schwarzschild spacetime (it is assumed that the radiating sphere does not contribute to the spacetime curvature). But the problem is not only interesting for the reason of the mathematical completeness of this solution. Such a~solution is particularly well suited to describe the interactions between black hole and the Cosmic Background Radiation, which corresponds to the uniformly luminous sphere located at infinity. The extended solution is also physically relevant when we consider the radiation from numerous, randomly spread sources, that can be approximated by a~homogeneous luminosity. That can be the case of a~black hole located in the center of a~spherical galaxy, for instance, or the radiatively efficient spherical accretion on the black hole.

% With the `natbib' package, two citation commands are distinguished:
% \citep is the paranthetical form, with complete citation surrounded by
% round brackets, while \citet is the textual form, with brackets enclosing
% the year only, often prepended by a `by', `in' etc, and being part of
% the sentence (see below).  The documentation of the `natbib' package
% treats this topic in full details:

\section{Case of the luminous inner sphere}\label{eom}%%%%%%%%%%%%%%%%%

\begin{figure}[t]
\begin{center}
\includegraphics[width=\linewidth]{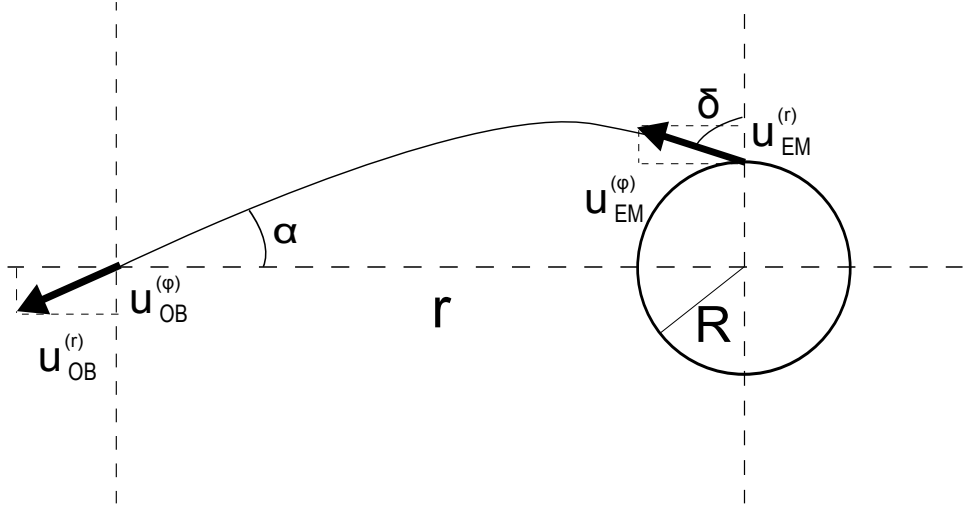}
\end{center}
\par\vspace{-1ex}\par
\caption{\label{Fig:Angles}Photon emitted from the star surface R with the emission angle $\delta$.}
\end{figure}

Let us first review the solution for an observer located above the static luminous sphere in Schwarzschild spacetime, given by \citet{Abramowicz1990}. We follow a~slightly different approach, giving some more general formulas, in order to make a very smooth extension to the case of $r < R$ (observer located at radius $r$ inside the luminous sphere of radius $R$) in the next section. Assuming homogeneous, isotropic radiation flux in the emitter's rest frame we conclude that the luminosity observed by any static observer located at a given radial location is constant over the observed sphere surface. Thus, the problem of calculating the radiation stress-energy tensor components in the static ZAMO frame reduces to the calculation of the constant specific intensity moments. Hence, radial dependence of the following two quantities need to be established
\begin{enumerate}
\item specific intensity $I(r)$ (value to be integrated over the observer's local sky),
\item sphere viewing angle $\alpha_0$ (boundary for the specific intensity integration).
\end{enumerate}
Specific intensity as seen by the observer located at radial distance $r$ corresponds to the surface intensity $I(R)$ of the radiation source, gravitationally redshifted by the presence of the central mass. The quantities that are conserved along the particular light ray traveling through the curved spacetime are the \emph{photon energy} 
\begin{equation}
 E = p^\alpha \eta_\alpha =0 = \mbox{const.}\, ,                                         \label{eq:PhotonEnergy}
\end{equation}
for the photon four-momentum $p^\alpha$ and Killing vector $\eta^\alpha = \delta^\alpha_{\ t}$, and the \emph{redshifted intensity}
\begin{equation}
  I_0 = \frac{I(r)}{\left( p^\alpha v_\alpha \right)^4} = \mbox{const.}\, ,                                        \label{eq:RedshiftedIntensity}
\end{equation}
for the stationary observer's four-velocity $v^\alpha = \eta^\alpha (|g_{tt}|)^{-1/2}$, see \citet{Misner73} section 4.22 for some more details. Hence, we find
\begin{equation}
I(r) = I(R)\left( \frac{1-2M/R}{1-2M/r} \right)^2 \, ,
\label{eq:Intensity}
\end{equation}
so clearly $I(r) < I(R)$ for $r > R$. Viewing angle $\alpha_0$ corresponds to the largest possible value of angle $\alpha$ in the Fig. \ref{Fig:Angles}, which occurs for the largest possible emission angle $\delta_0$ for which the photon can be observed. For $R \geq 3M$ $\delta_0 = \pi/2$, i.e., all emitted photons are able to escape from the star vicinity. This is not true for $2M < R < 3M$, where
\be
\sin^2 \delta_0 = 27(1-2M/R)/R^2 \, .
\label{Eq:Sind0}
\ee 
This result will be explained a little further.
The relevant angles can be defined using the photon four-velocity in the local orthonormal frame $u^{(\alpha)}$, i.e.,
\be
\tan \alpha = \left[ \frac{u^{(\varphi)}}{u^{(r)}}\right]_{OB}  = \left[ \frac{g_{\varphi \varphi}(u^{\varphi})^2}{g_{rr}(u^{r})^2}\right]^{1/2}_{OB} \ \ ; \ \ \tan \delta = \left[ \frac{u^{(\varphi)}}{u^{(r)}}\right]_{EM}  =  \left[ \frac{g_{\varphi\varphi}(u^\varphi)^2}{g_{rr} (u^r)^2} \right]^{1/2}_{EM} \, ,
\label{Eq:Angles}
\ee
lower subscripts OB and EM denoting the location of the photon emission and observation, respectively. Using the normalization $u^\alpha u_\alpha = 0$ it is easy to show that the following equation \emph{always} holds
\be
\frac{r^2 \sin^2\gamma }{1-2M/r} = - \frac{u_\varphi}{u_t} = \ell = \mbox{const.} \, ,
\label{Eq:SinIdentity}
\ee
where $\tan \gamma = u^{(\varphi)}/u^{(r)}$, $\gamma_{OB} = \alpha$ and $\gamma_{EM} = \delta$, from which we find that
\be 
\sin^2 \alpha_0 = \left(\frac{R}{r}\right)^2 \left( \frac{1-2M/r}{1-2M/R} \right) \sin^2 \delta_0 \, 
\ee
and since $\alpha_0(r)$ must decrease with radius, the \emph{only} solution is
\be
\alpha_0(r) = \left\{
  \begin{array}{l l}
     \arcsin\left[\frac{R}{r}\left(\frac{1-2M/r}{1-2M/R} \right)^{1/2} \right] & \quad \text{for $ 3M \leq R \leq r$}\\
     \arcsin\left[\frac{3\sqrt{3}M(1-2M/r)^{1/2}}{r} \right]& \quad \text{for $R < 3M< r$} \\
   \pi - \arcsin\left[\frac{3\sqrt{3}M(1-2M/r)^{1/2}}{r} \right]& \quad \text{for $R < r < 3M$ }
  \end{array} \right.
  \label{Eq:AlphaInner}
\ee
Finally, having calculated the $I(r)$ and $\alpha_0(r)$ distributions, the stress-energy tensor ZAMO components are found by the integration over the observer's local sky
\be 
T^{(\alpha)(\beta)}(r) = I(r)\int n^{(\alpha)}n^{(\beta)} d \Omega  \ \ ; \ \ n^{(\alpha)} = p^{(\alpha)}/p^{(t)} \, 
\ee
to give
\be
T^{(t)(t)} = 2 \pi I(r) (1 - \cos \alpha_0 ) \ \text{,}
\label{eq:Ttt0}
\ee
\be
T^{(t)(r)} = \pi I(r) \sin^2 \alpha_0  \ \text{,}
\label{eq:Ttr0}
\ee
\be
T^{(r)(r)} = \frac{2}{3} \pi I(r) (1 - \cos^3 \alpha_0 )  \ \text{,}
\label{eq:Trr0}
\ee
\be
T^{(\theta)(\theta)} = T^{(\varphi)(\varphi)} = \frac{1}{3} \pi I(r) (2 -3 \cos \alpha_0 + \cos^3 \alpha_0 )  \ \text{.}
\label{eq:Tthth0}
\ee 
All other components are simply equal to zero.
\section{Luminous outer sphere}\label{Lab:OuterSphere}%%%%%%%%%%%%%%%%%%%%%%%%%%%%
\begin{figure}[b]
\begin{center}
\includegraphics[width=\linewidth]{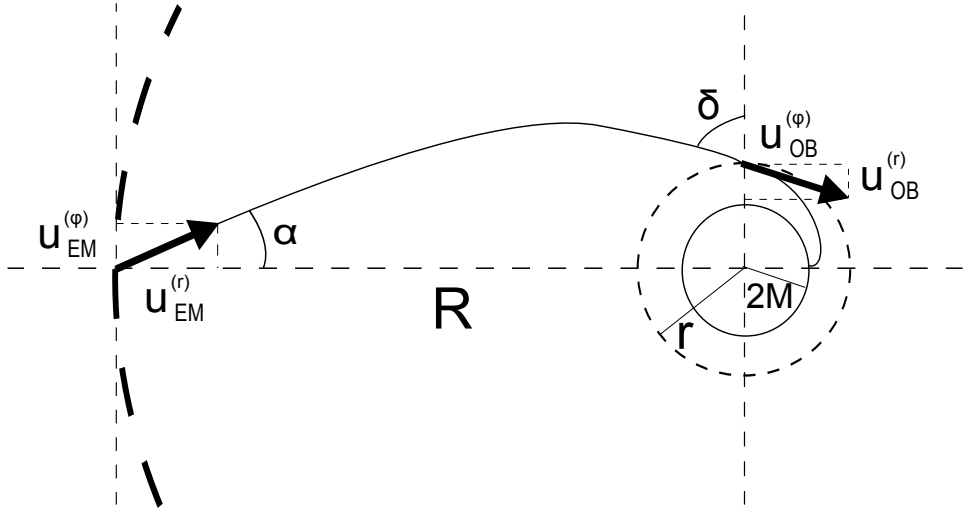}
\end{center}
\par\vspace{-1ex}\par
\caption{\label{Fig:AnglesOut}Photon emitted from the outer sphere surface R with the emission angle $\alpha$.}
\end{figure}

The case of the observer located inside the luminous sphere is similar to certain extent, since in the static spacetime photons may travel along the same null geodesic trajectory in both directions. Hence, the relation (\ref{eq:Intensity}) holds all the same, only difference being that $I(r) > I(R)$ for $R>r$, i.e., radiation is now \emph{blueshifted} in the static observer's frame. Equation (\ref{eq:Intensity}) also ensures that the observer's local sky is uniformly bright with exception of the part occluded by the black hole. Hence, the remaining part is to calculate the angular diameter of the black hole as a function of radius. For the analogy with the previous case, we denote this quantity with $2 \alpha_0$, see Fig. \ref{Fig:AnglesOut}. Note that it follows from the Eq. (\ref{Eq:SinIdentity}) that photons \emph{always} cross the horizon $r = 2M$ with angle $\gamma = 0$, i.e., perpendicularly to the horizon surface.
Let us now investigate the relation (\ref{Eq:SinIdentity}) in more details. One may notice that for a~photon trajectory to extend from $r_0 \gg 2M$ to the black hole horizon it is necessary that we are able to define a~meaningful photon radial four-velocity component for every $r_0 > r > 2M$, i.e., if
\be
u^r u_r > 0 \implies u_\varphi u^\varphi + u_t u^t < 0 \implies \ell^2 \leq - \frac{g_{\varphi\varphi}}{g_{tt}} \leq 27 \, .
\ee
27 is a~value of a~global maximum of $-g_{\varphi \varphi}/g_{tt}$ that occurs for $r = 3M$. This means that only photons with $\ell < 3 \sqrt{3}$ fall into the black hole and putting the maximum value of $\ell$ into Eq. (\ref{Eq:SinIdentity}), we find Eq. (\ref{Eq:Sind0}) (remember that the outgoing trajectory in the case of inner luminous sphere corresponds to the ingoing trajectory in the outer luminous sphere case). Considering that $\alpha_0$ must decrease monotonously with $r$, the \emph{only} solution for $\alpha_0(r)$ that satisfies Eq. (\ref{Eq:SinIdentity}) is
\be
\alpha_0(r) = \left\{
  \begin{array}{l l}
     \arcsin\left[\frac{3\sqrt{3}M(1-2M/r)^{1/2}}{r} \right] & \quad \text{for $r \geq 3M$}\\
   \pi - \arcsin\left[\frac{3\sqrt{3}M(1-2M/r)^{1/2}}{r} \right]& \quad \text{for $r < 3M$ }
  \end{array} \right.
\ee
which is the same as the Eq. (\ref{Eq:AlphaInner}) in the case of inner radiating sphere radius $R < 3M$. Note that the formula for $\alpha(r)$ does not depend on luminous sphere radius $R$. Finally, after the local sky integration, we find
\be
T^{(t)(t)} = 2 \pi I(r) (1 + \cos \alpha_0) \ \text{,}
\label{eq:Ttt}
\ee
\be
T^{(t)(r)} = - \pi I(r) \sin^2 \alpha_0  \ \text{,}
\label{eq:Ttr}
\ee
\be
T^{(r)(r)} = \frac{2}{3} \pi I(r) (1 + \cos^3 \alpha_0 )  \ \text{,}
\label{eq:Trr}
\ee
\be
T^{(\theta)(\theta)} = T^{(\phi)(\phi)} = \frac{1}{3} \pi I(r) (2 + 3\cos \alpha_0 - \cos^3 \alpha_0 )  \ \text{.}
\label{eq:Tthth}
\ee 
This system is quite similar to the result of \citet{Abramowicz1990}, yet the flux is of a different sign, and the angle is substituted $\alpha_0 \rightarrow \pi - \alpha_0$. It is interesting to observe, that  $\alpha_0 \rightarrow \pi$ as $r \rightarrow 2M$, so the area of integration (bright sky region) goes to zero as the observer approaches the horizon. On the other hand, $I(r)$ given by Eq. (\ref{eq:Intensity}) diverges in such a limit. So does the result of integration, the stress-energy tensor components, vanish in the limit of the horizon, diverge or have some finite limit? The answer to this question and its implications are discussed in details by \citet{Wielgus2014}.

\section{Conclusions}\label{conclus}%%%%%%%%%%%%%%%%%%%%%%%%%%%%%%%%%%%%%%%

We presented the extension of the classic analytic calculation of a~static luminous star radiation stress-energy tensor to the case of a~luminous sphere observed from the inside. We found out that because of the symmetries involved, such a~problem has a~very similar solution. Analysis of the Cosmic Background Radiation field properties close to the black hole horizon is one example of application of the presented formulas.

% Acknowledgements are created using the command \ack:
\ack%%%%%%%%%%%%%%%%%%%%%%%%%%%%%%%%%%%%%%%%%%%%%%%%%%%%%%%%%%%%%%%%%%%%%%%

% Contributors involved in `Vyzkumny zamer' can use macro \InstResCode
% instead of specifying the alphanumerical code explicitly:
This work was supported by the Czech Grant
CZ.1.07/2.3.00/20.0071 (“Synergy”, Opava)  as well as the Polish NCN grant UMO-2011/01/B/ST9/05439. We thank George F. R. Ellis and Frederic Vincent for many illuminating discussions about the subject of radiation treatment in general relativity.

% Here we specify the basename of the bibliography database file,
% in this case \jobname=ragsamp:
\bibliography{RAG_Wielgus}

\begin{thebibliography}{8}
\expandafter\ifx\csname natexlab\endcsname\relax\def\natexlab#1{#1}\fi
\expandafter\ifx\csname url\endcsname\relax
  \def\url#1{\texttt{#1}}\fi
\expandafter\ifx\csname urlprefix\endcsname\relax\def\urlprefix{URL }\fi
\providecommand{\selectlanguage}[1]{\relax}
\providecommand{\eprint}[2][]{\url{#2}}

\bibitem[{{Abramowicz} et~al.(1990){Abramowicz}, {Ellis} and
  {Lanza}}]{Abramowicz1990}
{Abramowicz}, M.~A., {Ellis}, G.~F.~R. and {Lanza}, A. (1990), {Relativistic
  effects in superluminal jets and neutron star winds}, \emph{Astrophysical
  Journal}, \textbf{361}, pp. 470--482.

\bibitem[{{Bini} and {Geralico}(2010)}]{Bini2010}
{Bini}, D. and {Geralico}, A. (2010), {Spinning bodies and the
  Poynting-Robertson effect in the Schwarzschild spacetime}, \emph{Classical
  and Quantum Gravity}, \textbf{27}(18), 185014, \eprint{1107.2793}.

\bibitem[{Misner et~al.(1973)Misner, Thorne and Wheeler}]{Misner73}
Misner, C.~W., Thorne, K.~S. and Wheeler, J.~A. (1973), \emph{{Gravitation}},
  Freeman, San Francisco.

\bibitem[{{Sok Oh} et~al.(2011){Sok Oh}, {Kim} and {Mok Lee}}]{Oh2011}
{Sok Oh}, J., {Kim}, H. and {Mok Lee}, H. (2011), {Finite size effects on the
  Poynting-Robertson effect: A fully general relativistic treatment}, \emph{New
  Astronomy}, \textbf{16}, pp. 183--186, \eprint{1011.3104}.

\bibitem[{{Stahl} et~al.(2013){Stahl}, {Klu{\'z}niak}, {Wielgus} and
  {Abramowicz}}]{Stahl2013}
{Stahl}, A., {Klu{\'z}niak}, W., {Wielgus}, M. and {Abramowicz}, M. (2013),
  {Escape, capture, and levitation of matter in Eddington outbursts},
  \emph{Astronomy \& Astrophysics}, \textbf{555}, A114, \eprint{1306.6556}.

\bibitem[{{Stahl} et~al.(2012){Stahl}, {Wielgus}, {Abramowicz}, {Klu{\'z}niak}
  and {Yu}}]{Stahl2012}
{Stahl}, A., {Wielgus}, M., {Abramowicz}, M., {Klu{\'z}niak}, W. and {Yu}, W.
  (2012), {Eddington capture sphere around luminous stars}, \emph{Astronomy \&
  Astrophysics}, \textbf{546}, A54, \eprint{1208.2231}.

\bibitem[{{Wielgus} et~al.(2014){Wielgus}, {Ellis}, {Vincent} and
  {Abramowicz}}]{Wielgus2014}
{Wielgus}, M., {Ellis}, G.~F.~R., {Vincent}, F.~H. and {Abramowicz}, M.~A.
  (2014), {Cosmic background radiation in the vicinity of a Schwarzschild black
  hole: No classic firewall}, \emph{Physical Review D}, \textbf{90}(12),
  124024, \eprint{1406.6551}.

\bibitem[{{Wielgus} et~al.(2012){Wielgus}, {Stahl}, {Abramowicz} and
  {Klu{\'z}niak}}]{Wielgus2012}
{Wielgus}, M., {Stahl}, A., {Abramowicz}, M. and {Klu{\'z}niak}, W. (2012),
  {Oscillations of the Eddington capture sphere}, \emph{Astronomy \&
  Astrophysics}, \textbf{545}, A123, \eprint{1208.2939}.

\end{thebibliography}

\end{document}